\documentclass[epj]{svjour}
\usepackage{graphics}
\usepackage{enumitem}
\usepackage{amssymb,amsmath}
\usepackage{graphicx}
\usepackage{cite}
\usepackage{xcolor}
\newcommand{\z}[1] 		{{\ensuremath{z_{#1}}}}
\newcommand{\rsj} 		{{\ensuremath{R_{\mathrm{sj}}}}}
\newcommand{\qhat} 		{\ensuremath{\hat{q}}}
\newcommand{\dsj} 		{\ensuremath{\Delta S_{12}}}

\newcommand{\mvr} 		{{\it med/vac}}
\newcommand{\pythia}    {\textsc{Pythia}}
\newcommand{\qpythia}   {\textsc{Q-Pythia}}
\newcommand{\pyquen}    {\textsc{PyQuen}}
\newcommand{\jewel}    	{\textsc{Jewel}}
\newcommand{\pqm}    	{\textsc{PQM}}
\newcommand{\sdrop}    	{\textsc{Soft Drop}}
\newcommand{\fastjet}   {\textsc{FastJet}}
\newcommand{\pp}           {pp}

\newcommand{\gevc}         {\ensuremath{\mathrm{GeV}/c}}

\newcommand{\pt}           {\ensuremath{p_{\mathrm{T}}}}
\newcommand{\ptjet}        {\ensuremath{p_{\mathrm{T}}^{\mathrm{jet}}}}

\newcommand{\kT}           {\ensuremath{k_{\rm T}}}
\newcommand{\akT}           {\ensuremath{{\rm anti-}k_{\rm T}}}
\newcommand{\tn}[1]{\textnormal{#1}}
\newcommand{\kt}{\ensuremath{k_\tn{T}}}

\begin{document}
\title{Novel subjet observables for jet quenching in heavy-ion collisions}
\author{Liliana Apolin\'{a}rio\inst{1,2}\thanks{liliana@lip.pt} \and Jos\'{e} Guilherme Milhano\inst{1,2,3}\thanks{guilherme.milhano@ist.utl.pt} \and Mateusz Ploskon\inst{4}\thanks{mploskon@lbl.gov} \and Xiaoming Zhang\inst{5}\thanks{xiaoming.zhang@cern.ch}% etc
}                     % Do not remove
\institute{LIP, Av. Prof. Gama Pinto, 2, 1649-003 Lisbon, Portugal
\and IST University of Lisbon, Av. Rovisco Pais 1, 1049-001, Lisbon, Portugal
\and Theoretical Physics Department, CERN, CH-1211, Geneva 23, Switzerland
\and Lawrence Berkeley National Laboratory, Berkeley, California 94720, USA
\and Institute of Particle Physics, Central China Normal University, Wuhan 430079, China}
\date{Received: date / Revised version: date}
\abstract{
Using a novel observable that relies on the momentum difference of the two most energetic subjets within a jet \dsj\ we study the internal structure of high-energy jets simulated by several Monte Carlo event generators that implement the partonic energy-loss in a dense partonic medium.
Based on inclusive jet and dijet production we demonstrate that \dsj\ is an effective tool to discriminate between different models of jet modifications over a broad kinematic range.
The new quantity, while preserving the collinear and infrared safety of modern jet algorithms, it is experimentally attractive because of its inherent resilience against backgrounds of heavy-ion collisions.
 % end of PACS codes
} %end of abstract
\maketitle
\section{Introduction}
\label{sec:intro}

Interactions of high-energy partons with a strongly coupled hot partonic medium - a quark-gluon plasma (QGP) \cite{Satz:2000bn,Bass:1998vz,Shuryak:1984nq,Cleymans:1985wb} - created in heavy-ion collisions, leading to modifications of the internal jet structure (jet quenching), was first proposed in \cite{Bjorken:1982tu} and is studied as a sensitive probe of the medium properties \cite{Wiedemann:2009sh,Burke:2013yra, Mehtar-Tani:2013pia}.
Experiments at RHIC and the LHC observed a strong suppression of high transverse momentum particle yields \cite{Adams:2005dq,Adcox:2004mh,Arsene:2004fa,Back:2004je, Aamodt:2010jd, Aamodt:2011vg, CMS:2012aa}, suppression of inclusive and semi-inclusive yields of fully reconstructed jets \cite{Aad:2010bu, Chatrchyan:2011sx, Adam:2015ewa, Adamczyk:2017yhe,Adam:2015doa}, and, more recently, the internal structure of the jets \cite{Aad:2014wha,Chatrchyan:2013kwa,Chatrchyan:2012gw,Sirunyan:2017bsd,Cunqueiro:2015dmx} for detailed studies of jet quenching.
However, in all these measurements the treatment of the background originating from the copiously produced particles not associated to hard scatterings poses an experimental challenge for precise and unbiased measurements.
Previous works that addressed the effect of filtering on subjet analysis \cite{Rubin:2010fc} and the recent analytic calculations of the momentum distributions of subjets \cite{Kang:2017mda,Larkoski:2015lea,Larkoski:2017bvj} and groomed jet mass distributions \cite{Frye:2016aiz,Marzani:2017mva,Marzani:2017kqd} that were recently measured in proton-proton collisions \cite{CMS:2017tdn,Aaboud:2017qwh}, provide a strong motivation for novel studies.
In this writeup, following previous works \cite{Zhang:2015trf}, we propose observables that are sensitive to the internal jet structure but significantly alleviate the difficulties associated to the effects of the background. Our approach is attractive from the experimental point of view and obeys the theoretical requirements of the infrared and collinear safety.

While jet substructure techniques are extensively used in the high-energy $pp$ collisions \cite{Chatrchyan:2013vbb,ATLAS:2012am,Tripathee:2017ybi} the first studies in the heavy-ion context \cite{Sirunyan:2017bsd,Kauder:2017cvz} are recent (see \cite{Zhang:2015trf} for a first attempt of using subjets as a phenomenological tool for jet quenching studies).
In this manuscript we propose an observable that uses only the highest and next-to-highest energetic fully reconstructed subjets within a jet.
This choice aims to minimize the impact of the heavy-ion background on the extracted jet properties allowing for better experimental control.

\section{Observable definition and setup}

We introduce a new jet substructure observable $\Delta S_{12}$ defined as the difference between the fractions of transverse momentum of a jet carried by its leading (hardest) and subleading (second hardest) subjets. That is
\begin{equation}
	\Delta S_{12} = z_{1} - z_{2}\, ,
	\label{eq:deltas12}
\end{equation}
where
\begin{equation}
	z_{i} = p_{{\rm T},i} / p_{{\rm T}, {\rm jet}}\, .
	\label{eq:zi}
\end{equation}
The subjets used to evaluate eq.~(\ref{eq:deltas12}) are obtained as follows:
\begin{enumerate}
	\item For each event, reconstruct jets with the anti-\kt\ algorithm  \cite{Cacciari:2008gp} provided by the \fastjet\ package \cite{Cacciari:2011ma} with radius $R$ and within pseudo-rapidity $|\eta_{jet}| < \eta_{max}$;
	\item Within each jet, find subjets by reclustering the jet components with a smaller radius parameter $\rsj < R$. Retain the two hardest (highest-\pt) subjets.
\end{enumerate}

The subjet samples used in this study were obtained with $ \eta_{max}=2.5$ and $R=0.5$.
In general, the reclustering of the jet components into subjets in step (ii) above can be carried out with a different jet algorithm from that chosen to reconstruct the jet to which they belong. We chose to use  anti-\kt\ after assessing the discriminating power of $\Delta S_{12}$ for subjets reconstructed with different algorithms and checking its sensitivity to hadronization effects (see Subsections \ref{sec:rsj_dep} and \ref{sec:Hadro})
The subjet radius parameter was set to $\rsj =0.15$ except when assessing, in  Subsection  \ref{sec:rsj_dep},  the dependence of $\Delta S_{12}$ on \rsj\ (where the range $0.1<\rsj <0.2 $ was considered), and when comparing, in Subsection \ref{sec:zg_vs_dsj}, with the analysis \cite{Sirunyan:2017bsd} (where we used $\rsj=0.1$).

The bulk of soft particles produced in high-energy collisions is not a priori distinguishable from the particles produced from the hadronisation of an energetic parton shower.
The presence of these background particles is the main experimental confounding factor when establishing the jet energy scale and jet energy resolution in  jet quenching studies (see for example \cite{Abelev:2012ej}).
Moreover, unlike in measurements of proton-proton collisions with high event pile-up probability within the detectors, the background in heavy-ion collisions is complex.
It consists of region to region fluctuations, modified particle production as compared to \pp\ collisions, and particle correlations caused by the collective expansion of the QGP.
In consequence, experimental observables at relatively low jet energies at the LHC ($\pt < 150$ GeV) are prone to systematic uncertainties related to complicated multi-dimensional unfolding procedures that are susceptible to large correction factors.

The substructure observable \dsj\ defined in eq.~\eqref{eq:deltas12} has been constructed to minimize correlated background contributions. Take \cite{Cacciari:2007fd}
\begin{equation}
  \Delta S_{12} = \frac{p_{T,1}^{\mathrm true}-p_{T,2}^{\mathrm true}}{p_{T,{\mathrm jet}}} = \frac{(p_{T,1}^{\mathrm rec} - \rho_{1} A_{1}) - (p_{T,2}^{\mathrm rec} - \rho_{2} A_{2})}{p_{T,{\mathrm jet}}},
 \label{eq:s12}
\end{equation}
where $p_{T,i}^{\mathrm true}$ is the true subjet momentum, $A_{i}$ is the area of a subjet, $\rho_{i}$ is the level of noise corresponding to the amount of transverse momentum added to each subjet per unit area by the background, and the $p_{T,i}^{\mathrm rec}$ is the experimentally reconstructed subjet momentum containing the background contribution $\rho_{i}A_{i}$.
For subjets reconstructed with the same radius parameter \rsj\ (in our case $\rsj = 0.15$) with the anti-\kt\ algorithm, the corresponding active areas are necessarily very similar $A_{1} \simeq A_{2}$. In an ideal case, where  $\rho_{1}=\rho_{2}$, the background term in the numerator of \dsj\ vanishes. For real events, where subjets sit close by, $\rho_{1}$ and $\rho_{2}$ can only differ by very localized fluctuations and thus should be on average still very similar. Thus, the background effect in the numerator of eq.~\eqref{eq:deltas12} is small.

A variety of observables similar to \dsj\ can be defined. In particular, \z{i}\ in eq.~\eqref{eq:zi} could be redefined by replacing the denominator  by the sum of the momenta of the leading and subleading subjets, such that $z_{i} = p_{{\rm T},i}/(p_{{\rm T},1} + p_{{\rm T},2})$.
Although such a definition could  have some welcome consequences in reducing the influence of background effects in the reconstructed jet transverse momentum $p_{T,\mathrm{jet}}$ (denominator of eq. \ref{eq:zi}), all information on the overall hardness of the jet fragmentation, that is the fraction of jet momentum carried by the two hardest subjets, would be neglected.

\section{Models}

To assess the potential of the proposed observable we consider a set of Monte Carlo event generators which rely on different implementations of jet quenching.
This allows both for a comparison between theoretical calculations that is not limited by systematic uncertainties of the putative experimental measurement and to assess the potential of the observable as a discriminant of different modelling scenarios.

Below we provide a short description of each event generator considered in this study --- \qpythia\ v1.0.2 \cite{Armesto:2009fj}, \jewel\ v2.0.0 \cite{Zapp:2013vla}, and \pyquen\ v1.5.1 \cite{Lokhtin:2005px} --- emphasising only the main characteristics and details of the setup we adopted (for further details please see the corresponding references). All samples used in this work were generated for central (0-10\% most central) PbPb collisions at  $\sqrt{s}_{NN} = 2.76$ TeV.

\qpythia\ is a modification of  \pythia\ 6.4 \cite{Sjostrand:2006za} where the splitting probability in the final state parton shower is enhanced by an additional term that follows the BDMPS-Z radiation spectrum \cite{Baier:1996sk}.
The medium is modelled by a single parameter, a local in space and time transport coefficient $\hat{q}$ that translates the averaged transverse momentum squared $\left<q^{2}_{\rm T}\right>$ exchanged between a parton and the medium per mean free path $\lambda$ in that medium, such that $\hat{q}=\left< q^{2}_{T}\right>/\lambda$.  The time and spatial variation of  $\hat{q}$ is modelled following the  \pqm\ prescription \cite{Dainese:2004te}. We considered two different average $\hat{q}$ values ($\hat{q} = 1$ GeV$^2$ fm$^{-1}$ and $\hat{q} = 4$ GeV$^2$ fm$^{-1}$) known to capture the main jet quenching features observed in dijets \cite{Apolinario:2012cg}.

\jewel\ implements a description of jet evolution that takes into account both elastic and inelastic energy losses as all scatterings with the medium are described by infra-red continued leading order matrix elements for $2 \rightarrow 2$ processes. Additional medium-induced radiation is also taken into account during the jet development and can be induced by several coherent scatterings, as predicted by the LPM effect \cite{Landau:1953um,Migdal:1956tc}.
We kept all default settings and used the medium implementation with Bjorken expansion described in detail in \cite{Zapp:2013zya} validated on a large set of jet quenching observables \cite{Zapp:2013vla}.

\pyquen\ is a modification  (afterburner) of standard \pythia\ 6.4 jet events in which both radiative and collisional accumulated energy losses are applied during the parton shower development. The former is calculated for an expanding medium within the BDMPS framework, where the angular distribution follows three simple parameterisations (small, wide and collinear angular distributions) that are used for comparison purposes. The latter is calculated in the high-momentum transfer approximation. Additional in-medium gluon radiation is added at the end of the parton shower, before hadronization. We chose the internal parameters that characterise the QGP formation expected for central PbPb collisions at the LHC.

While a typical Monte Carlo reference for jet production in \pp\ collisions is constructed with \pythia\ \cite{Sjostrand:2006za} each of the models provides their own implementation and/or modifications of \pythia\ original routines and consequently their own \pp\ reference.
Therefore, when comparing the medium-modified jets with jets showering in vacuum we take the model-provided proton-proton collision equivalent.

\section{Results}

We provide examples of how \dsj\ can be used to discriminate between the different implementations of jet quenching (Sec. \ref{sec:deltaS}) and evaluate its sensitivity to the choice of algorithm for subjet clustering and subjet radius \rsj\ (Sec. \ref{sec:rsj_dep}).
Section \ref{sec:Dijets} illustrates how \dsj\ combined with a dijet analysis can be used to study jet quenching more differentially as compared to the inclusive measurements.
In section \ref{sec:Hadro} we show the robustness of the results against hadronization effects.
Finally, Sec. \ref{sec:zg_vs_dsj} provides an overview of the relation between \dsj\ and the recently explored $z_{g}$ observable in vacuum.

\subsection{\dsj\ as a model discriminant}
\label{sec:deltaS}

\begin{figure}[h]
  \centering
	\includegraphics[width=0.45\textwidth, page=8]{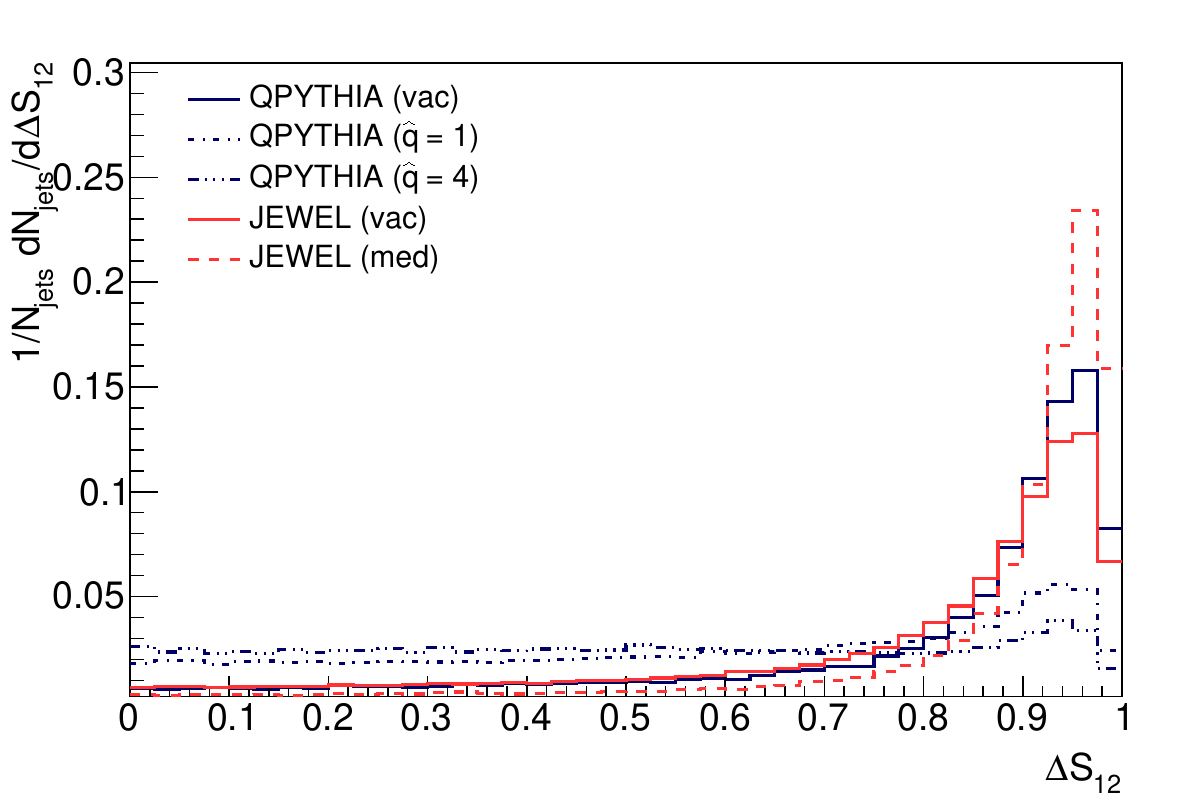}
	\includegraphics[width=0.45\textwidth, page=7]{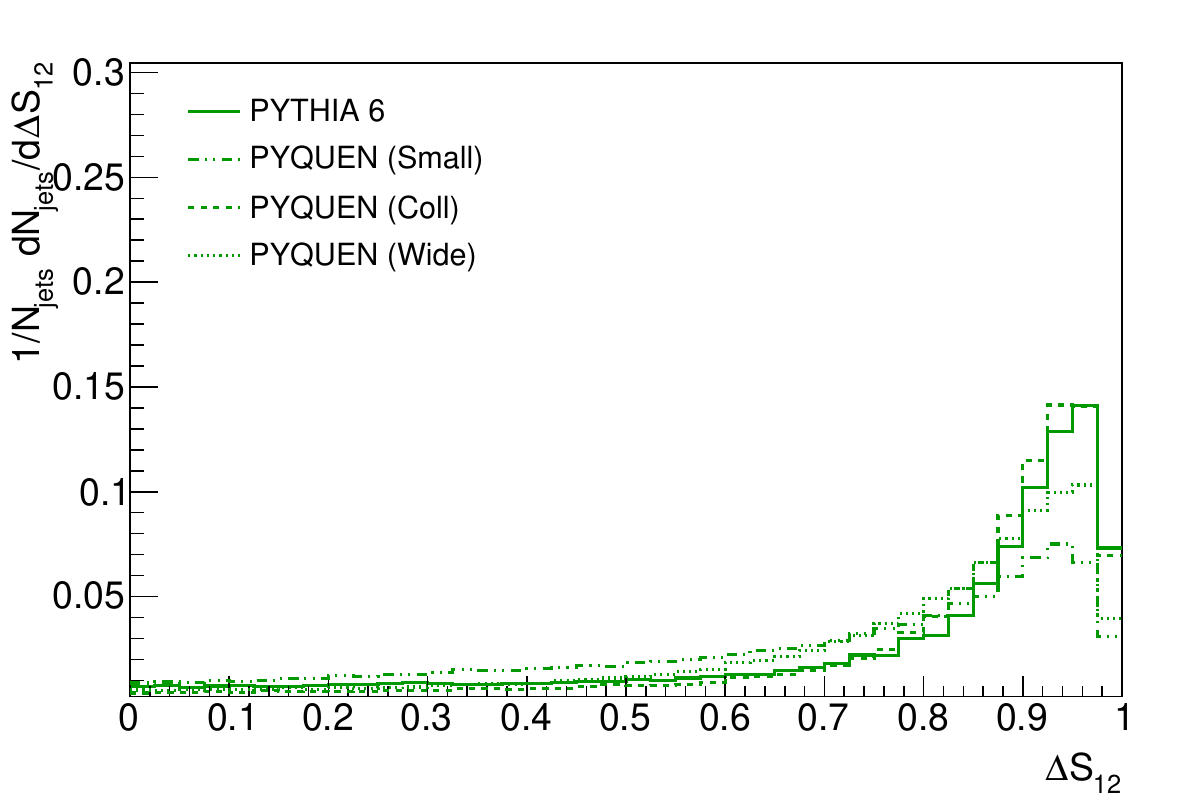}
	\caption{Distribution of \dsj\ for $\rsj=0.15$ \akT\ subjets within $R=0.5$ \akT\ jets with $\pt > 150$~\gevc. }
  \label{fig:DeltaS12}
\end{figure}

The distribution of the difference \dsj\ between the fractions of the jet total transverse momentum carried by the leading and subleading subjets is shown in Figure \ref{fig:DeltaS12}.

In vacuum -- \qpythia\ (vac) and \jewel\ (vac) (top panel), and \pythia\ 6 (bottom panel) -- the distribution displays a pronounced maximum for $\dsj>0.9$ and a tail towards lower \dsj\ values. Medium effects in \qpythia\ and \jewel\ (top panel) modify the \dsj\ distribution in incompatible directions.  \jewel\   enhances significantly the maximum of the distribution and mildly depletes its tail.  \qpythia\ softens the peak at high \dsj\ and produces a flat tail towards values of \dsj\ $\leq 0.7$ with the effects more pronounced for increasing $\hat{q}$. These observations are consistent with a collimation of jets in \jewel\ and broadening in \qpythia\ as compared to their vacuum references. \pyquen\ (Coll) (bottom panel) modifies the \dsj\ similarly to \jewel, \pyquen\ (Small) gives a distribution with features resembling those found for \qpythia,  and \pyquen\ (Wide) displays an intermediate behaviour.

While the above features directly reflect the behaviour of the \z{1}\ and \z{2}\ distributions (see appendix \ref{app: zdist}), we emphasise that, from the experimental point of view, studies of \dsj\ are more attractive as compared to the individual \z{i} distributions since the difference $\z{1}-\z{2}$ removes, by construction, a large fraction of the correlated background. Although the effect of the uncorrelated background is enhanced in \dsj\ ratio with respect to \z{2}\ on a jet-by-jet basis, we found that this effect is small when taking the integrated/inclusive distributions, and subsequently their moments. Moreover, \dsj\ is more robust for low momentum jets for which \z{2} becomes gradually (with decreasing \pt) dominated by background particles.
u

\vspace{5mm}
To further expose the differences among models we now turn our focus to the quartiles of the \dsj\ distribution, considering  \mvr\ ratios $\mathcal{R}$, where {\it med} refers to calculations including jet quenching effects and \textit{vac} to the corresponding model specific no-quenching baseline
\begin{equation}
	\mathcal{R}_{\mathsf{Q_i} [\dsj]} = \frac{\mathsf{Q_i}[\dsj]^{med}}{\mathsf{Q_i}[\dsj]^{vac}}\, .
\end{equation}

While the \mvr\ ratio of the medians of the \dsj\ distributions as a function of \ptjet\  display a clear evolution and discrimination power among the models (data not shown), we find that to characterise the modifications to the subjet structure for models that show jet collimation  -- \jewel\ and \pyquen\ (Coll) -- the ratios of the first quartile ($\mathsf{Q_1}$) of \dsj\ distributions is preferable. The criteria of selecting the \textit{best} discriminant was made by calculating the relative standard deviation (RSD) given by the models in each $\pt$ bin, i.e, the ratio of the standard deviation over the mean. The relative spread among the different models for each observable is thus quantified (a larger spread translates into a larger RSD) and it can be used as a guiding parameter to select the observable that maximizes the differences among jet quenching models.

\begin{figure}[h]
  \centering
	\includegraphics[width=0.5\textwidth, page=1]{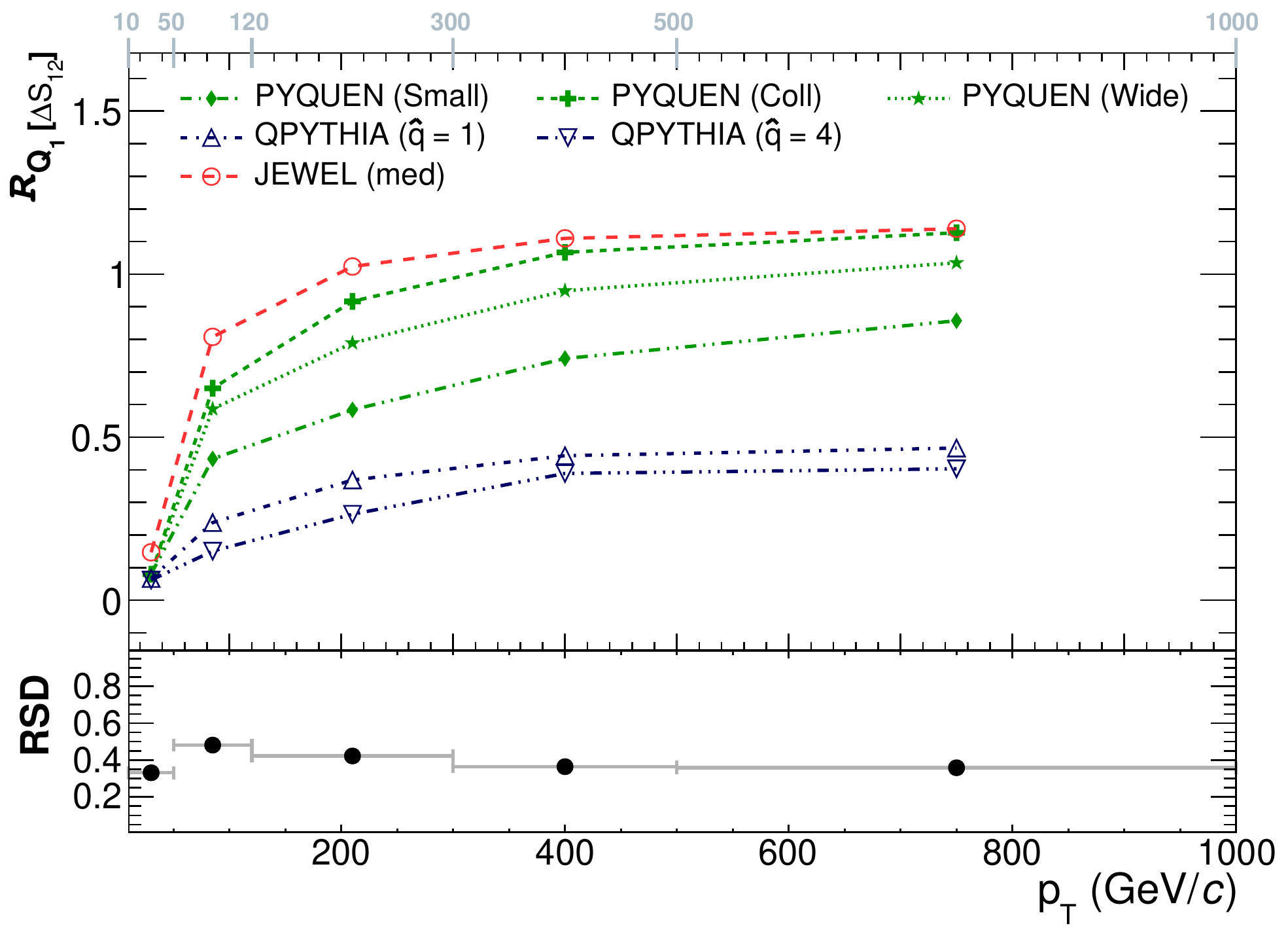}
  \caption{\textit{Top panel}: $\mathcal{R}_{\mathsf{Q_1} [\dsj]}$ as a function of \ptjet; \textit{Bottom panel}: RSD\ of $\mathcal{R}_{\mathsf{Q_1} [\dsj]}$ as a function of \ptjet. The edges of the five considered  jet \pt\ bins  (10-50, 50-120, 120-300, 300-500, 500-1000 \gevc) are shown on the top of the figure.}
  \label{fig:mvr_Q1_dsj}
\end{figure}

The ratios $\mathcal{R}_{\mathsf{Q_1} [\dsj]}$ are shown in Figure \ref{fig:mvr_Q1_dsj}, in the upper panel, while the corresponding RSD\ in the bottom panel of the same figure.

Here we find a clear evolution with the jet momentum for jets with $\pt < 300$~\gevc\ \, for all models. Up to this jet $\pt$ all models show a suppression of $\mathcal{R}_{\mathsf{Q_1} [\dsj]}$ reflecting more balanced momentum sharing between the two leading subjet structures than in the vacuum references. However, at high jet $\pt$, this observable remains fairly constant and shows a strong sensitivity to models that produce jets with a more symmetric structure, such as \qpythia\ and \pyquen(Small) ($\mathcal{R}_{\mathsf{Q_1} [\dsj]} < 1$) separating them well apart from \pyquen\ (Wide and Coll) and \jewel\ ($\mathcal{R}_{\mathsf{Q_1} [\dsj]} \simeq 1$).

\begin{figure}[h]
  \centering
	\includegraphics[width=0.5\textwidth, page=4]{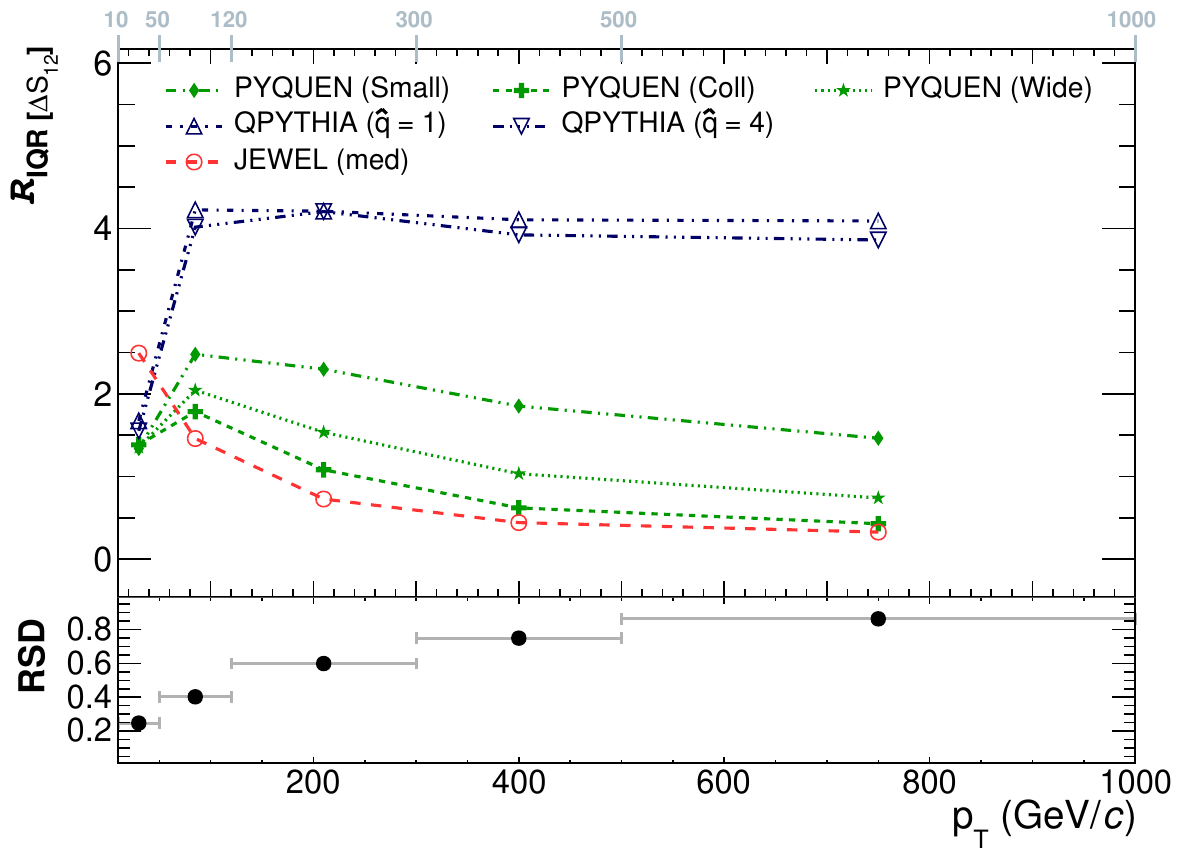}
  \caption{\textit{Top panel:} $\mathcal{R}_{\mathsf{IQR} [\dsj]}$ as a function of \ptjet; \textit{Bottom panel:} RSD\ of $\mathcal{R}_{\mathsf{IQR} [\dsj]}$ as a function of \ptjet. The edges of the five considered  jet \pt\ bins  (10-50, 50-120, 120-300, 300-500, 500-1000 \gevc) are shown on the top of the figure.}
  \label{fig:mvr_w_dsj}
\end{figure}

Further, we find that the interquartile range $\mathsf{IQR} = \mathsf{Q_3} -\mathsf{Q_1}$, that characterises the \textit{width} of the \dsj\ distribution, gives additional information. Figure \ref{fig:mvr_w_dsj} shows $\mathcal{R}_{\mathsf{IQR} [\dsj]}$ as a function of jet transverse momentum for the different quenching models with the corresponding RSD\ calculated in each $\pt$ bin.
Here again, models that result in jet collimation, characterised by a similar or narrower \dsj\ distribution than its vacuum reference ($\mathcal{R}_{\mathsf{IQR} [\dsj]} $ $\leq 1$), are clearly separated from those that broaden the jet, where \dsj\ is typically broader with respect to the vacuum reference ($\mathcal{R}_{\mathsf{IQR} [\dsj]} > 1$).

Moreover, for \qpythia, \jewel\ and \pyquen (Coll), $\mathcal{R}_{\mathsf{IQR} [\dsj]}$ converges quickly to a constant value with increasing jet $\pt$. Importantly, it also allows to better discriminate between the two models that destroy the vacuum subjet asymmetry: while \qpythia\ is well separated from its vacuum reference for all $\pt > 100$~\gevc\,, \pyquen (Small) evolves slowly towards more asymmetric jets with increasing \pt .
Thus, $\mathcal{R}_{\mathsf{IQR} [\dsj]}$ provides relevant complementary information to identify the main characteristics of jet quenching within specific models, in particular for jets with $100 < \pt < 200$~\gevc, where the first quartile $\mathsf{Q_1}$ of the \dsj\ is suppressed with respect to the vacuum reference in all models.
Comparing the RSD\ of $\mathcal{R}_{\mathsf{Q1} [\dsj]}$ and $\mathcal{R}_{\mathsf{IQR} [\dsj]}$ we find that for low $\pt$ jets (jets with $\pt < 120$~\gevc), the $\mathcal{R}_{\mathsf{Q1} [\dsj]}$ has a higher discrimination power while $\mathcal{R}_{\mathsf{IQR} [\dsj]}$ is preferable for higher $\pt$ jets. Nonetheless, it should be noted that $\mathsf{IQR}$ is also more sensitive to hadronization effects (see section \ref{sec:Hadro})

\vspace{5mm}
We also investigated the evolution of \dsj\ with the relative distance $\Delta R_{subjet}$ in $(\eta, \phi)$ space between the leading and subleading subjets, in particular of the \mvr\ of its median value $\mathsf{Q_2}$  and interquartile range $\mathsf{IQR}$. The corresponding RSD\ are calculated in each $\Delta R_{subjet}$ bin.

\begin{figure}[h]
  \centering
	\includegraphics[width=0.5\textwidth, page=2]{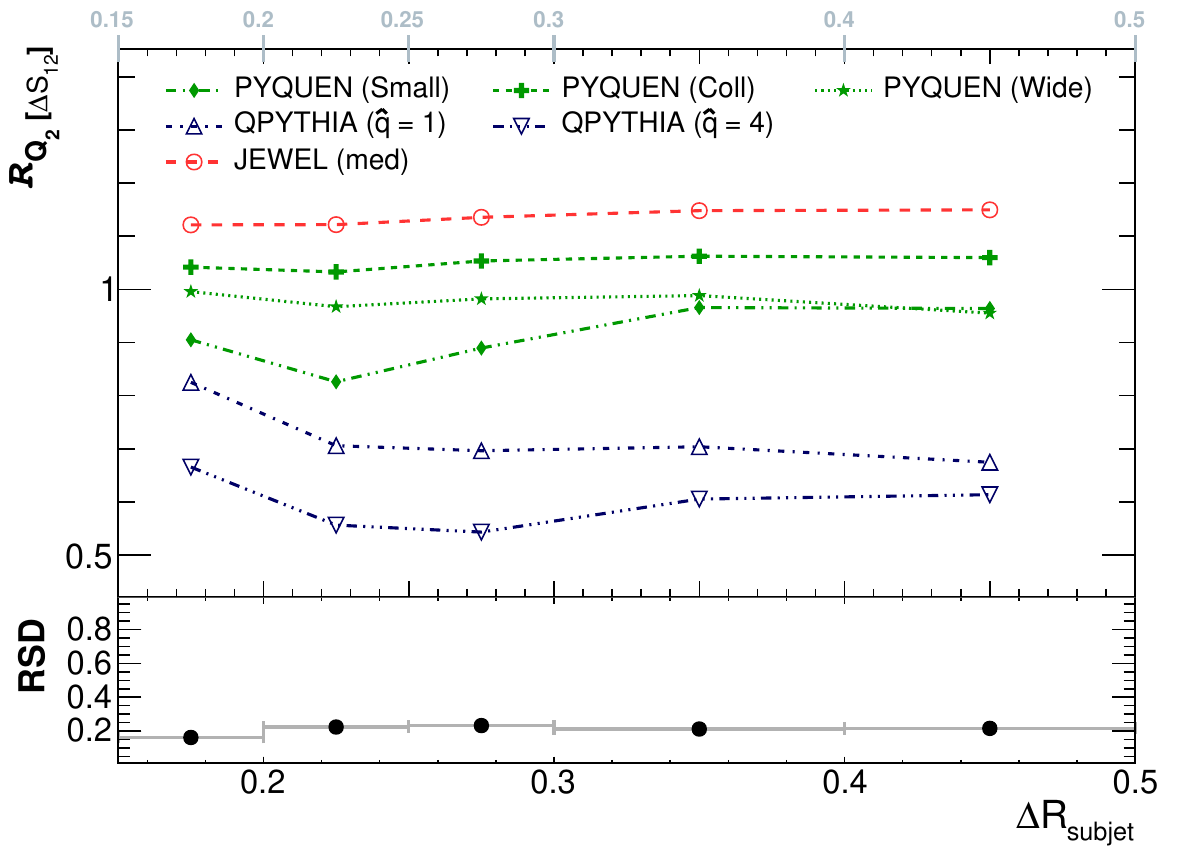}
  \caption{\textit{Top panel:} $\mathcal{R}_{\mathsf{Q_2} [\dsj]}$ as a function of $\Delta R_{subjet}$; \textit{Bottom panel:} RSD\ of $\mathcal{R}_{\mathsf{Q_2} [\dsj]}$ as a function of $\Delta R_{subjet}$. The edges of the five considered $\Delta R_{subjet}$ bins: (0.15-0.2; 0.2-0.25, 0.25-0.3; 0.3-0.4; 0.4-0.5) are shown on the top of the figure.}
  \label{fig:mvr_Q1_dsj2}
\end{figure}

\begin{figure}[h]
  \centering
	\includegraphics[width=0.5\textwidth, page=4]{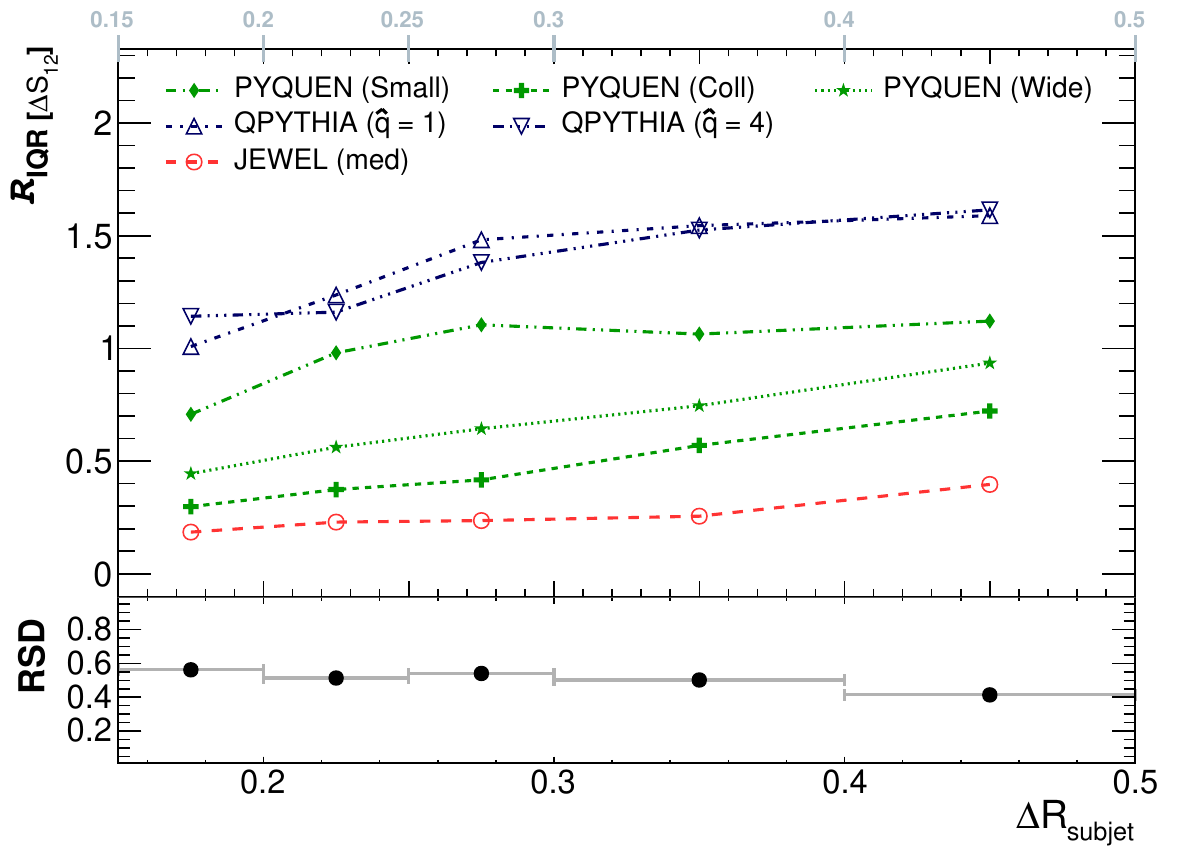}
  \caption{\textit{Top panel:} $\mathcal{R}_{\mathsf{IQR} [\dsj]}$ as a function of $\Delta R_{subjet}$; \textit{Bottom panel:} RSD\ of $\mathcal{R}_{\mathsf{IQR} [\dsj]}$ as a function of $\Delta R_{subjet}$. The edges of the five considered $\Delta R_{subjet}$ bins: (0.15-0.2; 0.2-0.25, 0.25-0.3; 0.3-0.4; 0.4-0.5) are shown on the top of the figure.}
  \label{fig:mvr_w_dsj2}
\end{figure}

The median ratio (Figure \ref{fig:mvr_Q1_dsj2}) shows a clear separation between models  --  \qpythia\ and \pyquen (Small) -- that broaden the jet structure. In \qpythia\ the two leading subjets become more symmetric with increasing $\Delta R_{subjet}$ (the ratio is below one and decreases). The same behaviour is seen for \pyquen (Small) up to $\Delta R_{subjet} = 0.25$, but interestingly, vacuum-like behaviour is recovered for larger separations. In contrast, the median ratio in \jewel\ and \pyquen (Coll) show a dependence on $\Delta R_{subjet}$ similar to their vacuum references with the ratio nearly independent of the distance between the two leading subjets. The large reduction of the interquartile range ($\mathcal{R}_{\mathsf{IQR} [\dsj]} < 0.5$) for all $\Delta R_{subjet}$ observed (Figure \ref{fig:mvr_w_dsj2}) in these models provides another clear signature of the jet collimation effect.

From the RSD\ values the interquartile ratio allows to have a wider spread between the models, although the transition from \pyquen (Small) to vacuum-like behaviour is more noticeable through the median ratio.

\subsection{Dependence on the choice of subjet clustering algorithm and subjet radius}
\label{sec:rsj_dep}

To investigate the dependence of the proposed observable and its sensitivity to the effects of jet quenching we varied the subjet reconstruction algorithm as well as the subjet radius parameter $\rsj<R$.

\begin{figure}[h]
  \centering
    \includegraphics[width=0.5\textwidth, page=2]{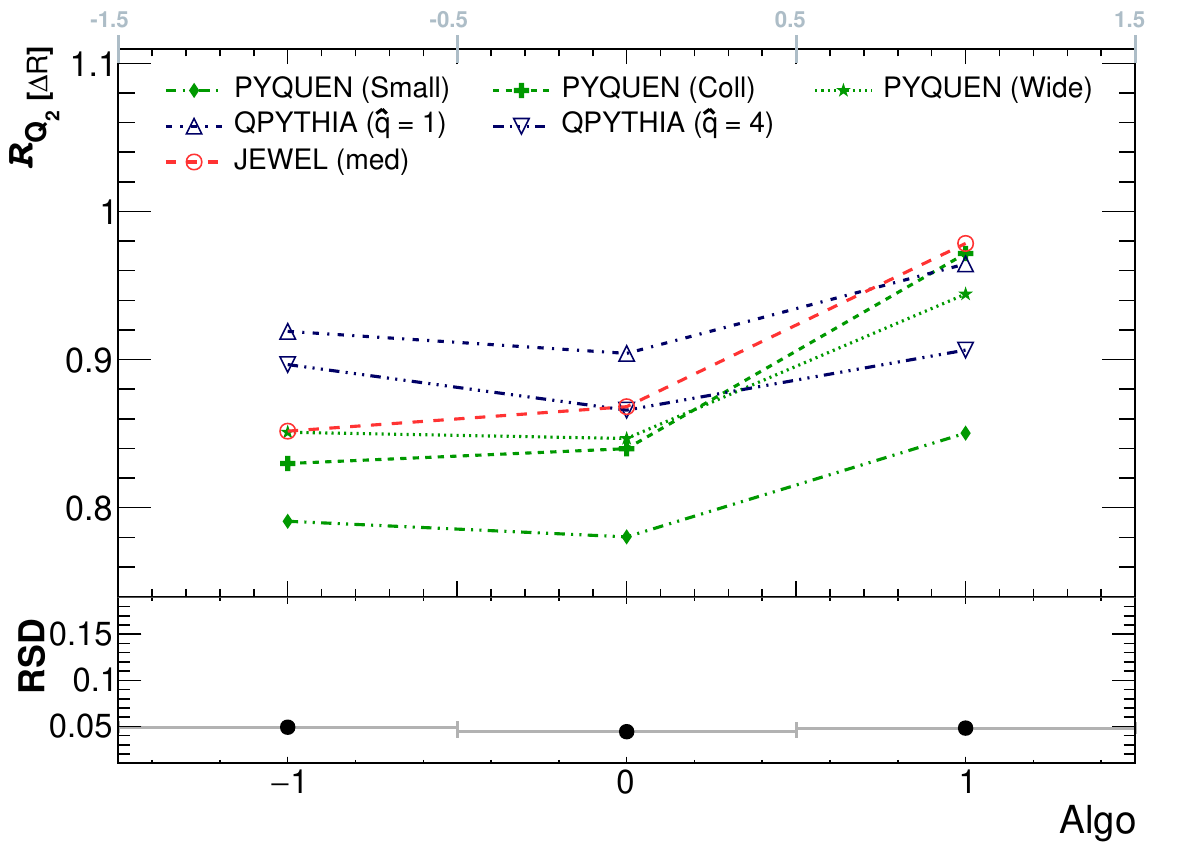}
  \caption{Evolution of the \mvr\ ratio of the medians of $\Delta R$ distributions $\mathcal{R}_{\mathsf{Q_2} [\Delta R]}$ (\textit{top panel}) and corresponding RSD\ (\textit{bottom panel}) with the subjet reconstruction algorithm with radius of $\rsj = 0.15$ for anti-\kT\ jets with $\pt > 150$~\gevc.}
  \label{fig:ds_algo}

\end{figure}

We find no significant differences in \dsj\ subjet distributions when changing the clustering algorithm from \akT\ \cite{Cacciari:2008gp} to \kT\ \cite{Ellis:1993tq} or Cambridge-Achen (C/A) \cite{Dokshitzer:1997in}. However, $\Delta R_{subjet}$ depends, by construction, on the reconstruction algorithm. Figure \ref{fig:ds_algo} shows the \mvr\ ratio of the medians of $\Delta R_{subjet}$ distributions for different models for $\pt>150$ ~\gevc\ jets with the subjet radius set, as before, to $\rsj=0.15$, and the corresponding RSD\ for completeness. The integer values -1, 0, 1 on the x-axis correspond, respectively, to \akT, C/A, and \kT. 
Despite the finite differences between clustering algorithms, we find that the power of discrimination between the different models is largely independent of the choice of the algorithm. This observation, together with the results obtained in section \ref{sec:Hadro}, where we study the effect of different hadronization models on the reconstructed subjets, allow us to conclude that the anti-\kt\ algorithm provides the most promising option when optimising for jet quenching effects. We therefore adopt this clustering algorithm as the standard setting for the remainder of this work.

\begin{figure}[h]
  \centering
    \includegraphics[width=0.5\textwidth, page=2]{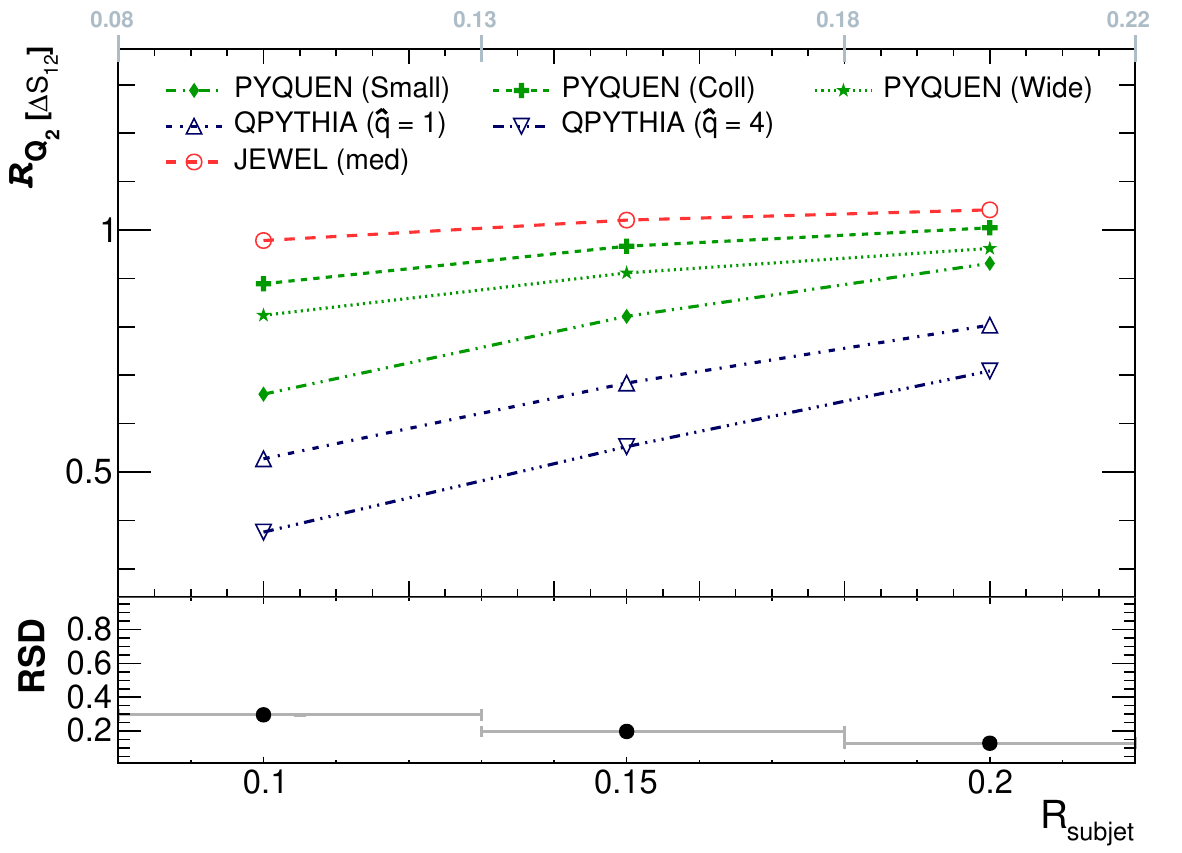}
  \caption{Evolution of the \mvr\ ratio of the medians of \dsj\ distributions $\mathcal{R}_{\mathsf{Q_2} [\dsj]}$ (\textit{top panel}) and corresponding RSD\ (\textit{bottom panel}) with the subjet radius \rsj\ for anti-\kT\ jets with $\pt > 150$~\gevc.}
  \label{fig:ds_rsub}
\end{figure}
Figure \ref{fig:ds_rsub} shows the dependence on $\rsj$ of the \mvr\ ratio of the medians of \dsj\ distributions for subjets reconstructed with the \akT\ algorithm. The RSD\ is also shown in the bottom panel, now calculated for each algorithm separately.
Here, we find that an increased discrimination between models resulting in jet collimation and the models preferring jet broadening is achieved with $\rsj \in [0.1; 0.15]$. Further, we find a clear difference in the energy distribution inside the jet that results from the different models. On the one hand, \qpythia\ and \pyquen (Small) increase the leading subjet $\pt$ when the subjet radius is increased, indicating broadening of the jet structure. On the other hand, in the models that produce collimated jets by medium effects, the energy in the leading subjet is nearly independent of the chosen subjet radius as it is highly concentrated close to the jet core.

\subsection{Subjets in dijet pairs}
\label{sec:Dijets}

\begin{figure}[h]
  \centering
  \includegraphics[width=0.5\textwidth, page=2]{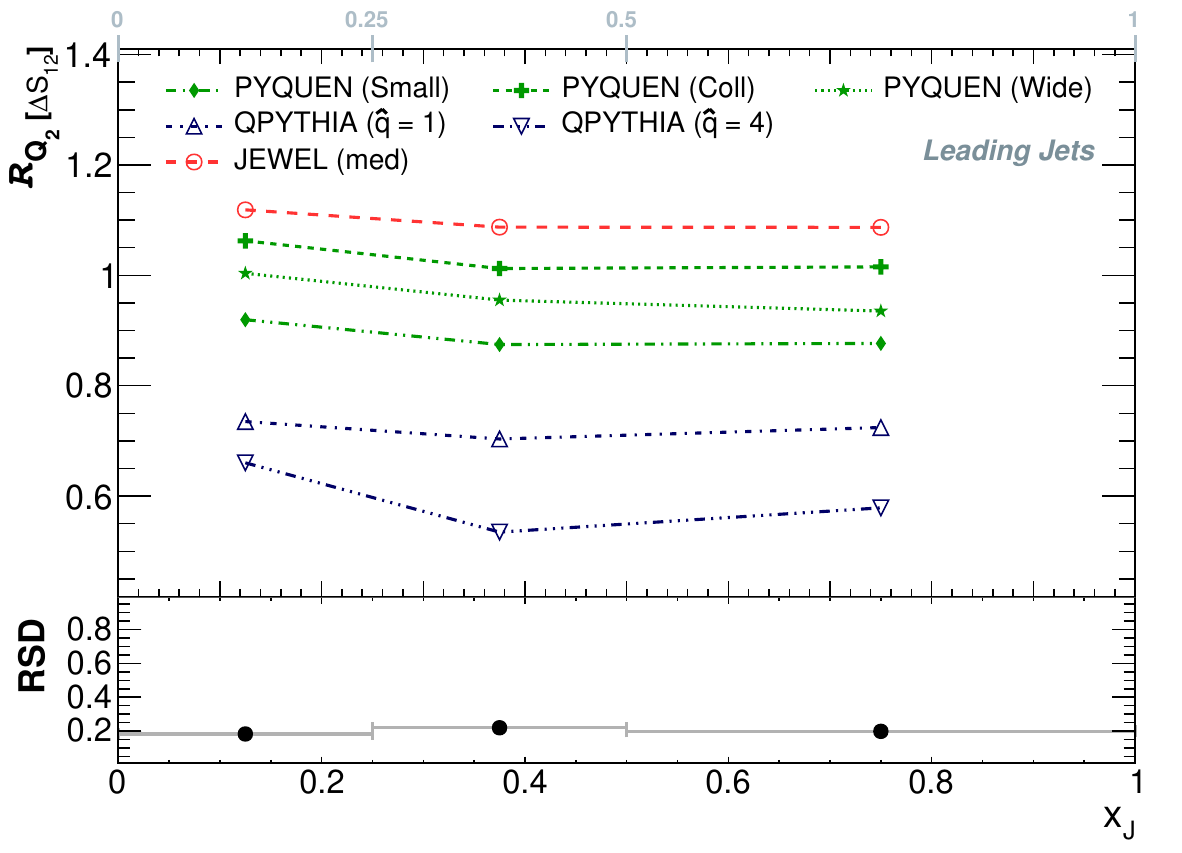}
  \includegraphics[width=0.5\textwidth, page=6]{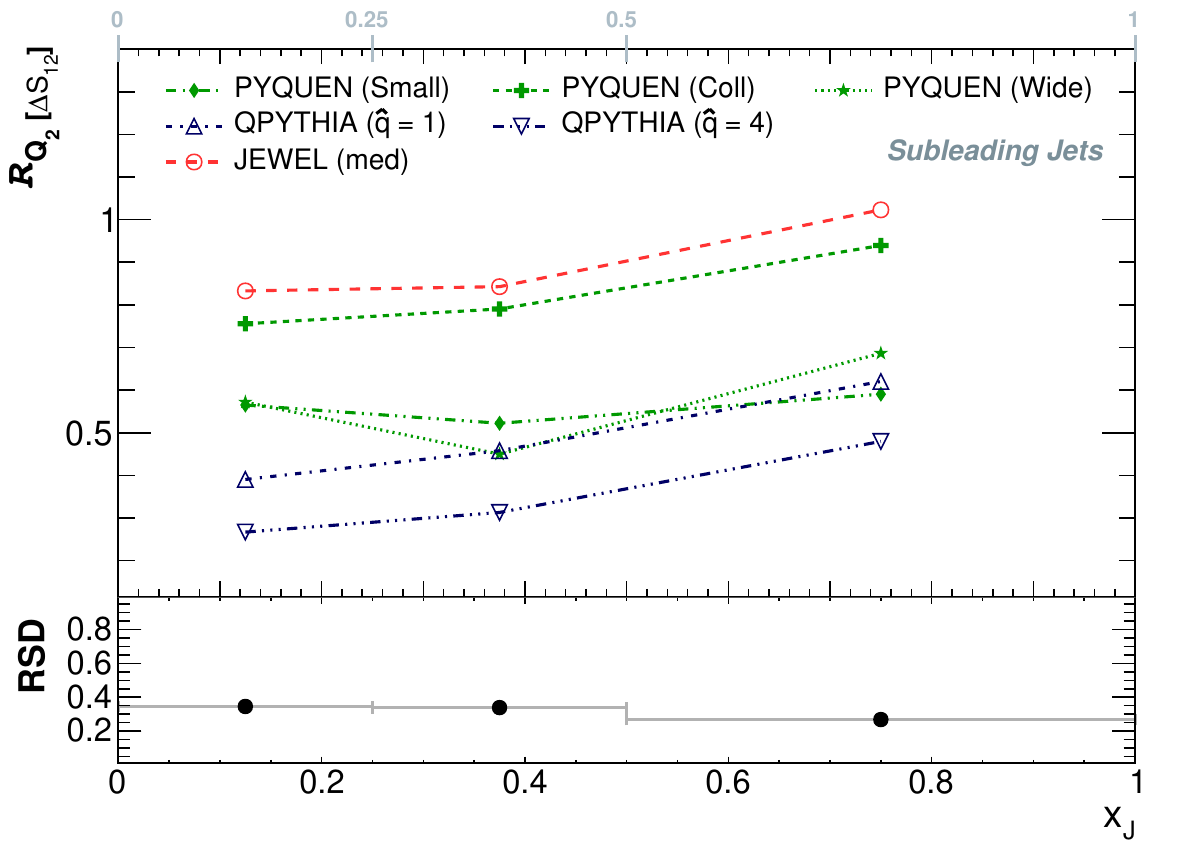}
  \caption{$\mathcal{R}_{\mathsf{Q_2} [\dsj]}$ as a function of $x_{J}$ for ({\it top figure}) leading jets and ({\it bottom figure}) recoil jets in dijet pairs. The edges of the three considered $x_J$ bins (0-0.25; 0.25-0.5; 0.5-1) are shown on the top of the figure. The bottom panel of each figure show the corresponding RSD.}
  \label{fig:mvr_median_dsj_xj}
\end{figure}

In a back-to-back dijet pair propagating through the QGP, the sub-leading jet has typically lost more energy than its leading partner \cite{Milhano:2015mng}. This quenching asymmetry can be combined with \dsj\ to experimentally further constrain the nature of jet quenching.
We have performed an analysis of dijet pairs with $R=0.5$ \akT\ jets within with $|\eta_{jet}| < 2$ where the leading jet was required to have $\pt>120$~\gevc\ and the recoil jet $\pt>50$~\gevc. The jets in the pair were required to be separated in azimuth by at least $5/6\pi$.
The \mvr\ ratios of medians of the \dsj\ distribution, $\mathcal{R}_{\mathsf{Q_2} [\dsj]}$, as a function of  $x_{J} = \pt^{\rm recoil~jet}/\pt^{\rm leading~jet}$ are shown in Figure \ref{fig:mvr_median_dsj_xj}. The upper figures show results for leading jets and the bottom figures those for recoil jets. Again, the upper panels of each figure show the evolution of all models and the bottom panels the corresponding spread quantified through the RSD\ calculated in each asymmetry bin.

Models that collimate jets towards their core -- \jewel\ and \pyquen\ (Coll) -- display leading jets with a (slightly) enhanced asymmetric subjet momentum balance as compared to their vacuum references, while the recoil jets have a more balanced subjet momentum distribution  than in vacuum.
In contrast, in models that broaden the jet structure -- \qpythia, \pyquen\ (Small), and to a more limited extent \pyquen\ (Wide) -- both leading and recoil jets have a more balanced subjet momentum distribution than in vacuum with the modification significantly stronger for recoil jets.
In all cases, the leading jet is modified independently of the pair asymmetry, while the momentum sharing between subjets becomes increasingly balanced (with respect to vacuum) with increasing dijet asymmetry (decreasing  $x_{J}$) for recoil jets.

\begin{figure}[h]
  \centering
  	\includegraphics[width=0.5\textwidth, page=4]{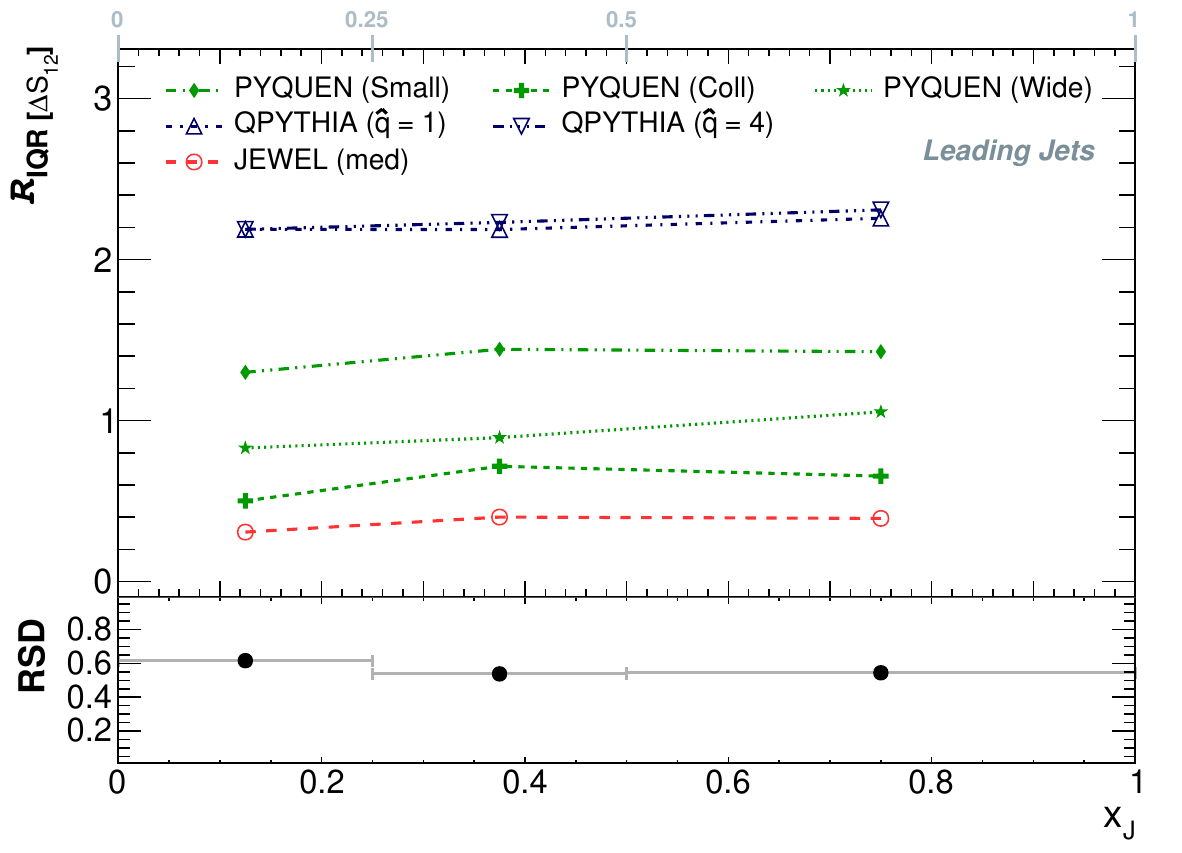}
  	\includegraphics[width=0.5\textwidth, page=8]{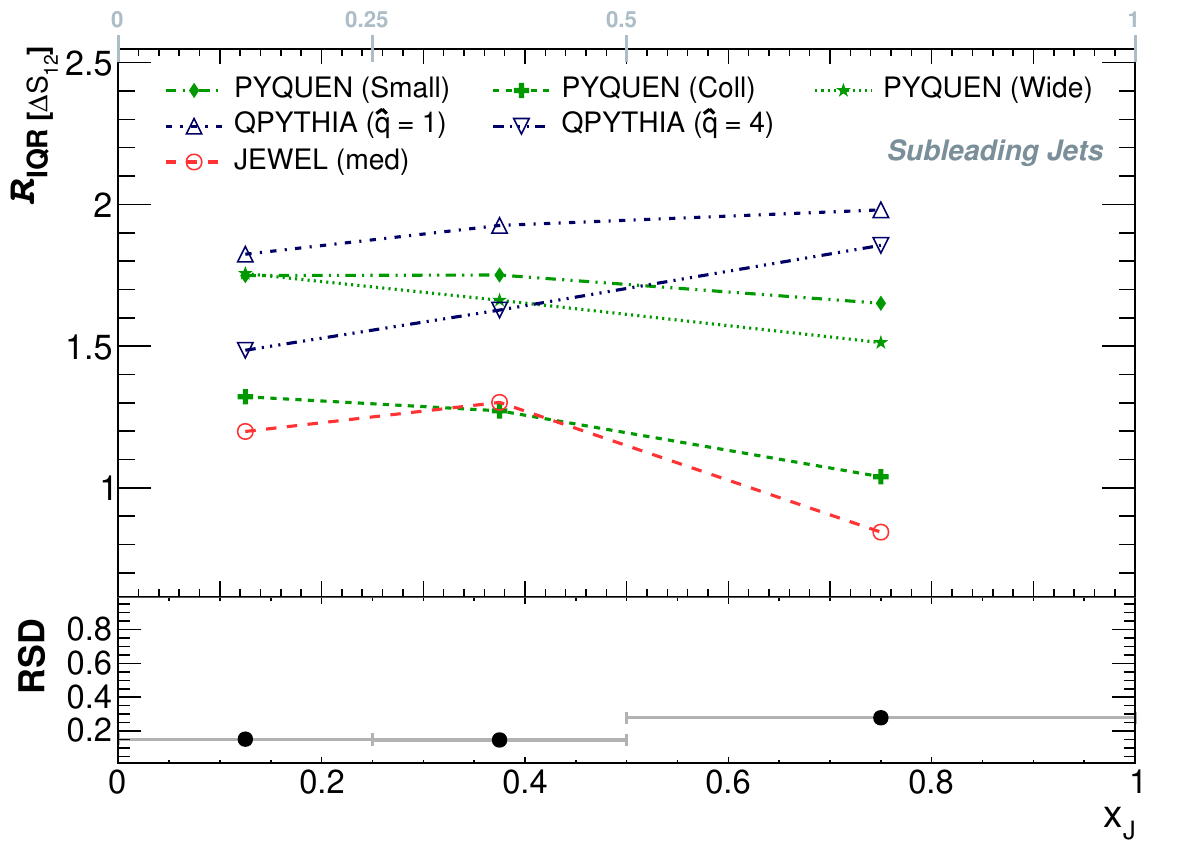}
  \caption{$\mathcal{R}_{\mathsf{IRQ} [\dsj]}$ as a function of $x_{J}$ for ({\it top figure}) leading jets and ({\it bottom figure}) recoil jets in dijet pairs. The edges of the three considered $x_J$ bins (0-0.25; 0.25-0.5; 0.5-1) are shown on the top of the figure. The bottom panel of each figure show the corresponding RSD.}
  \label{fig:mvr_irq_dsj_xj}
\end{figure}

Similar findings are also present in the ratio of the interquartile range of the distributions, for both leading and recoil jets, shown in Fig. \ref{fig:mvr_irq_dsj_xj}. However, differences between leading jets are more noticeable through the asymmetry of the \dsj\ distribution while recoil jets show a larger spread among models through $\mathcal{R}_{\mathsf{Q_2} [\dsj]}$

All these observations are consistent with the findings from section \ref{sec:deltaS}, figure \ref{fig:mvr_Q1_dsj}, where jets below $\pt = 200 \gevc$ always have a \dsj\ that is smaller than its vacuum reference. This is the preferred kinematic region for the recoil jet in unbalanced dijet systems. Moreover, $\mathcal{R}_{\mathsf{Q_1} [\dsj]}$ is fairly constant for jets above $\pt = 200$\gevc, where the leading jet (and recoil for balanced dijet systems) typically comes.

\subsection{Hadronization effects on the reconstructed subjets}
\label{sec:Hadro}

Small radii jets are known to be more sensitive to hadronization effects\cite{Dasgupta:2007wa}. For this reason, we investigate the role of different hadronization models in the distributions that were presented so far by using both PYTHIA 8 and HERWIG 7\cite{Bahr:2008pv,Bellm:2015jjp}. The former is based solely on the Lund string fragmentation framework \cite{Andersson:1983ia} while the later applies a cluster model\cite{Kupco:1998fx} to hadronize the resulting partonic final state to produce hadrons. Although such study is not ideal to accurately assess the uncertainties induced by hadronization effects, including in-medium hadronization modifications \cite{Beraudo:2011bh}, it can provide an estimate of the robustness of the proposed observable, \dsj.

We have found that $\mathsf{Q_2} [\dsj]$ is almost insensitive to the hadronization model, with relative differences (taking PYTHIA 8 as reference) smaller than $1\%$ for any choice of subjet radius or clustering algorithm and jets with a transverse momentum $p_{T,jet} > 100$~\gevc. For low momentum jets ($p_{T,jet} < 100$~\gevc), this difference goes up to $2\%$ for $R_{subjet} = 0.15$ and $10\%$ for $R_{subjet} = 0.1$.

As for the first quartile of the distribution, $\mathsf{Q_1} [\dsj]$, the relative change of HERWIG 7 with respect to PYTHIA 8 is $\sim [2-5]\%$ for anti-\kt\ subjets with $R_{subjet} \leq 0.15$. Any other choice of clustering algorithm or subjet radius provide a relative difference of $\sim [5-10]\%$ independently of the jet transverse momentum.

Finally, the interquartile range, $\mathsf{IQR} [\dsj]$ that is able to provide, in general, a larger dispersion between the jet quenching models, is also able to discriminate more among hadronization models. The relative change in low momentum jets ($p_{T,jet} < 100$~\gevc) between the interquartile range provided by the two Monte Carlo event generators is around $[4 -10]\%$ for any clustering algorithm and subjets reconstructed with $R_{subjet} \leq 0.15$. For $R_{subjet} = 0.2$, this change increases to $~17\%$, independently of the clustering algorithm. In high momentum jets ($p_{T,jet} > 250$~\gevc) the relative difference is around $[20 - 40]\%$. The lower bracketing is constantly observed for anti-\kt\ and small radius subjets while the upper bracketing occurs for \kt\ and large radius subjets. For the chosen parameters of this manuscript (anti-\kt\ subjets with $R_{subjet} = 0.15$), the relative change is $\sim 25\%$ for any jet with $p_{T,jet} > 100$\gevc.

The general large sensitivity of the interquartile range to the choice of the hadronization model comes from the fact that this observable is designed to promote the tails of the distributions. While it is the preferable region to tag energy loss modifications imprinted on the jet, it is also the region dominated by a fragmentation pattern that promotes the existence of one ($\dsj \sim 1$) or two ($\dsj \sim 2$) subjets mainly composed by very soft particles. Any modification on the hadronization mechanism would imply a stronger deviation on both $\mathsf{Q_1} [\dsj]$ (as observed from the increase of the relative differences with respect to $\mathsf{Q_2} [\dsj]$) and $\mathsf{Q_3} [\dsj]$.

These observations validate our choice of using reclustered anti-\kt\ subjets with $R_{subjet} = 0.15$ as to maximize jet quenching phenomena with respect to hadronization effects.

\subsection{Sub-jet momentum fraction $z_{g}$ and \dsj}
\label{sec:zg_vs_dsj}

\begin{figure}[h]
  \includegraphics[width=0.45\textwidth]{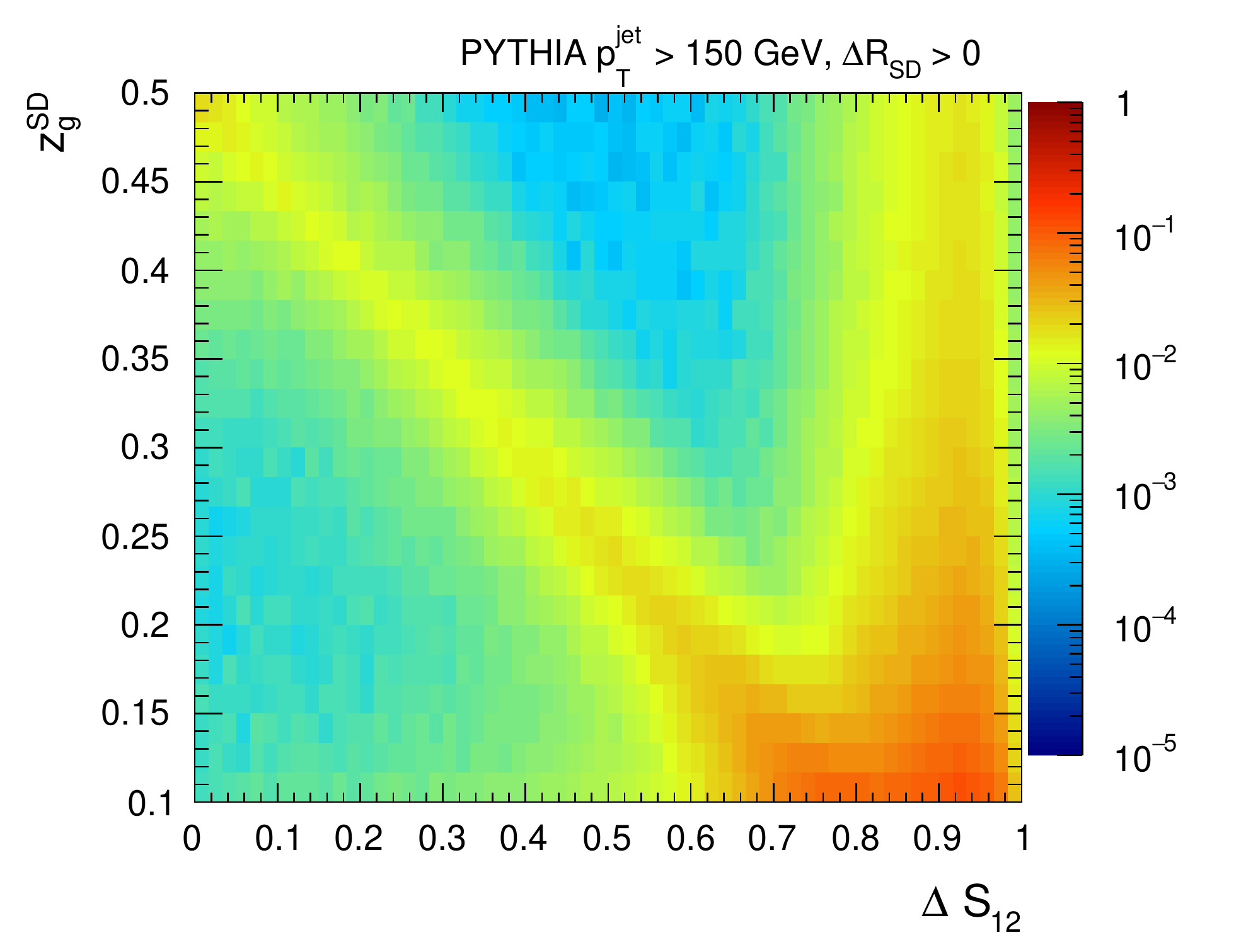}
  \includegraphics[width=0.45\textwidth]{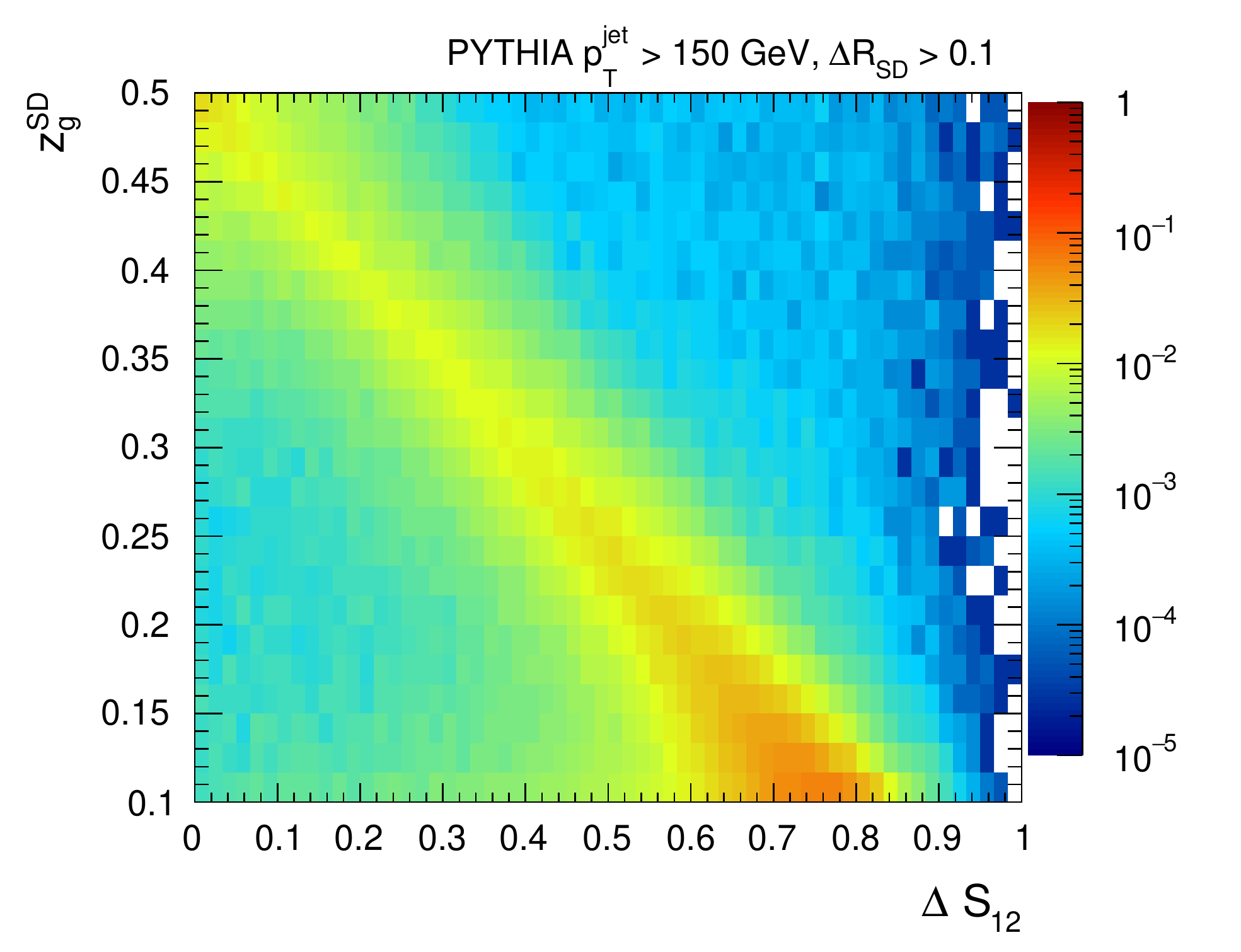}
  \caption{Subjet momentum fraction $z_{g}^{SD}$ reconstructed using the \sdrop\ algorithm as a function of \dsj. {\it Upper panel}: distribution for all jets with $\pt>150$ \gevc. {\it Lower panel}: distribution for jets for which subjets used for calculating $z_{g}$ are separated by a distance of $\Delta R > 0.1$.}
  \label{fig:zg_dsj}
\end{figure}

Recent studies of the momentum fraction $z_{g}$ in jets \cite{Sirunyan:2017bsd} prompt for a comparison of $z_{g}$ with \dsj. We have performed an analysis of (vacuum) \pythia\ jets with $\pt>150$~\gevc\ using settings of the \sdrop\ algorithm \cite{Dasgupta:2013ihk, Larkoski:2014wba} as in \cite{Sirunyan:2017bsd}. Figure \ref{fig:zg_dsj} shows the $z_{g}$ as a function of \dsj\ for two cases: one, where all jets where used; and a second, where jets with $\Delta R_{\rm sj} < 0.1$ between the subjets used to calculate $z_{g}$ are discarded. \dsj\ was calculated with $R_{subjet} = 0.1$ in both cases. We find a strong correlation between \dsj\ and the calculation of $z_{g}$ when using the $\Delta R_{\rm sj}$ cut as in \cite{Sirunyan:2017bsd}. Without the cut on $\Delta R_{\rm sj}$ the distributions have two dominating structures. One is the diagonal, but the other is largely independent of the \dsj\ at $\dsj>0.8$.

A comment on the differences of behaviour between these two observables in the presence of a QGP is in order. One of the proposed explanations \cite{Milhano:2017nzm} for the observed modification of the $z_g$ distribution in nucleus-nucleus collisions relies on the ability of subjets to collect contributions from the QGP. As argued in that work, this QGP backreaction process implies a distinctive increase in size (measured girth) of the subjets. Since in \dsj\ we explicitly impose, by specifying a radius parameter $\rsj$ for the subjet reconstruction, a size for the subjets, contributions from the QGP to both subjets will be of the same order, and thus, will cancel in \dsj . This makes \dsj\ and $z_g$, well correlated in vacuum, complementary observables in the presence of a QGP that can be used to disentangle the role of QGP backreaction from other dynamical processes that conceivably modify the jet substructure.

\section{Discussion}

\jewel. Results from \jewel\ are consistent with a \textit{jet collimation} effect, i.e., most of the radiation is transported outside of the cone leaving the energy core of the jet almost un-modified with respect to the vacuum reference but in a narrower region of phase space. We note, that in this analysis we have used \jewel\ in its "recoil-off'" mode which discards the medium partons that interacted with the jet. In this way, the results are independent of the medium-response, whose impact was recently analysed in \cite{Milhano:2017nzm}. As a consequence, the $\dsj$ is closer to unity for the medium modified jets and it does not change with the chosen subjet subjet radius. Moreover, the comparison of properties of the leading and subleading jets from a dijet event shows that \dsj\ increases for the leading jets and decreases for the (more strongly medium-modified/low momentum) subleading jet. This is an exclusive characteristic of the jet collimation phenomena and/or collisional energy loss as the same kind of behaviour is observed for \pyquen (Coll). In this model, since $\theta_{rad} = 0$, all the energy that is lost outside of the cone is due to elastic energy loss.

\qpythia. In \qpythia, which is as an implementation of the BDMPS-Z spectrum (without account for destructive interferences), the emission rate is enhanced according to the quenching parameter \qhat\ leading to a large modifications of the jet inner core. As a consequence, the distribution with a maximum for $\dsj > 0.9$ in vacuum shows a large tail to lower values due to in-medium interactions due to softening of the subjet spectrum (including the leading subjet). Such effect is visible for both leading and subleading jets. Moreover, the medium-induced gluon radiation is evenly distributed in phase space up to very large distances as \dsj\ mean value is constantly below the vacuum reference without a significant change for $\Delta R_{subjet} > 0.2$.

\pyquen. \pyquen\ considerations are centred around three angular distributions for the in-medium radiation spectrum. For \pyquen (Small) the finite angle of the radiation ($\theta < 5^{\circ}$) enhances the substructure and the impact on \dsj\ is qualitatively similar to \qpythia.  Nonetheless, a striking difference from this model with respect to \qpythia\ is the increasing asymmetry of the subjet structure when biasing the jet sample with $\Delta R_{subjets} > 0.3$. This could be due to the fact that since the radiation is displaced at a finite angle from the leading parton, the more the second hardest subjet is reconstructed away from the jet core, the less probable is to recover the energy. As such, the energy-momentum distribution inside of the jet is located at intermediate distances from the jet core in contrast to what happens in \qpythia .
A similar, but much milder modification of \dsj\ is observed for \pyquen (Wide). Since gluon radiation goes as $\sim 1/\theta$, the radiation is essentially kept near the core with few particles going to very large angles. This angular distribution is similar to the jet vacuum development, which makes this model undistinguishable from the vacuum reference for jets with a large transverse momentum and/or leading jets in dijet systems.
On the other hand, the \pyquen (Coll) mode, constrained to elastic energy loss only, affects mainly the softest jet constituents by such elastic collisions, whose energy is "absorbed" by the medium. As such, jets become more collimated, as it happens in \jewel.

Considerations of the RSD\ distributions show that the $\mathcal{R}_{\mathsf{IQR} [\dsj]}$ carries the largest discrimination power for high-$\pt$ jets, although there is an associated uncertainty of $[10-25]\%$ due to hadronization effects. For low-$\pt$ jets ($\pt \lessapprox 120~\gevc$) and/or recoil-jets in dijet systems the use of the $\mathcal{R}_{\mathsf{Q_1} [\dsj]}$ and/or the median $\mathcal{R}_{\mathsf{Q_2} [\dsj]}$ may prove more advantageous, with hadronization uncertainties that are smaller than $5\%$ for the chosen subjet parameters.

Finally, we reiterate that inclusive or semi-inclusive measurements of nuclear modification factor(s) for jets that fall within a range of \dsj\ (and $\Delta R_{subjets}$) can provide a rather straightforward insight into the properties of jet quenching (see \cite{Zhang:2015trf} for example).

\section{Conclusions}

We have presented observables of subjet structure that by minimizing the impact of the particle backgrounds in heavy-ion collisions are advantageous from the experimental point of view.
At the same time, the introduced \dsj\ quantity preserves the collinear and infrared safety of modern jet algorithms.
Using a number of Monte Carlo jet quenching models we have demonstrated that \dsj\ distribution and $\Delta R_{subjets}$ can be used as a sensitive tool to discriminate between different quenching mechanisms. We have shown that it is possible to use the quartiles of those distributions, together with the widths and/or use of dijets to make an accurate assessment of the main jet quenching characteristics, in particular, to determine the angular structure of the medium-induced gluon radiation and to investigate further the role of collisional energy loss in the in-medium shower development.

\section*{Acknowledgements}
The authors would like to thank to N. Armesto, M. Verweij and K. Zapp for useful discussions.

This work was supported in part by the U.S. Department of Energy, Office of Science, Office of Nuclear Physics, under contract DE-AC02-05CH11231 (MP, XZ) and by the Funda\c{c}\~ao para a Ci\^encia e Tecnologia (Portugal) under contracts CERN/FIS-NUC/0049/2015, Investigador FCT - Development Grant IF/00563/2012 (JGM) and SFRH/BPD/103196/2014 (LA).

\appendix

\section{Leading and subleading subjets}
\label{app: zdist}

\begin{figure}[h]
  \centering
  \includegraphics[width=0.45\textwidth, page=4]{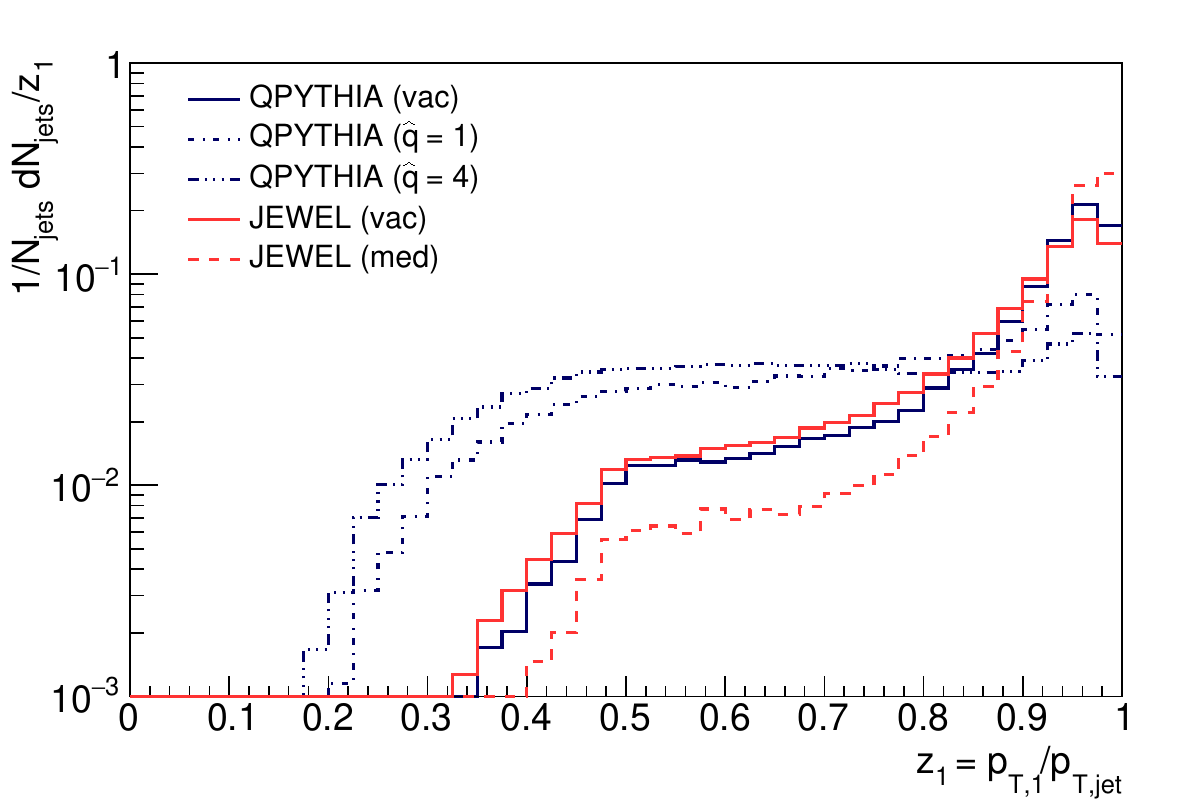}
  \includegraphics[width=0.45\textwidth, page=3]{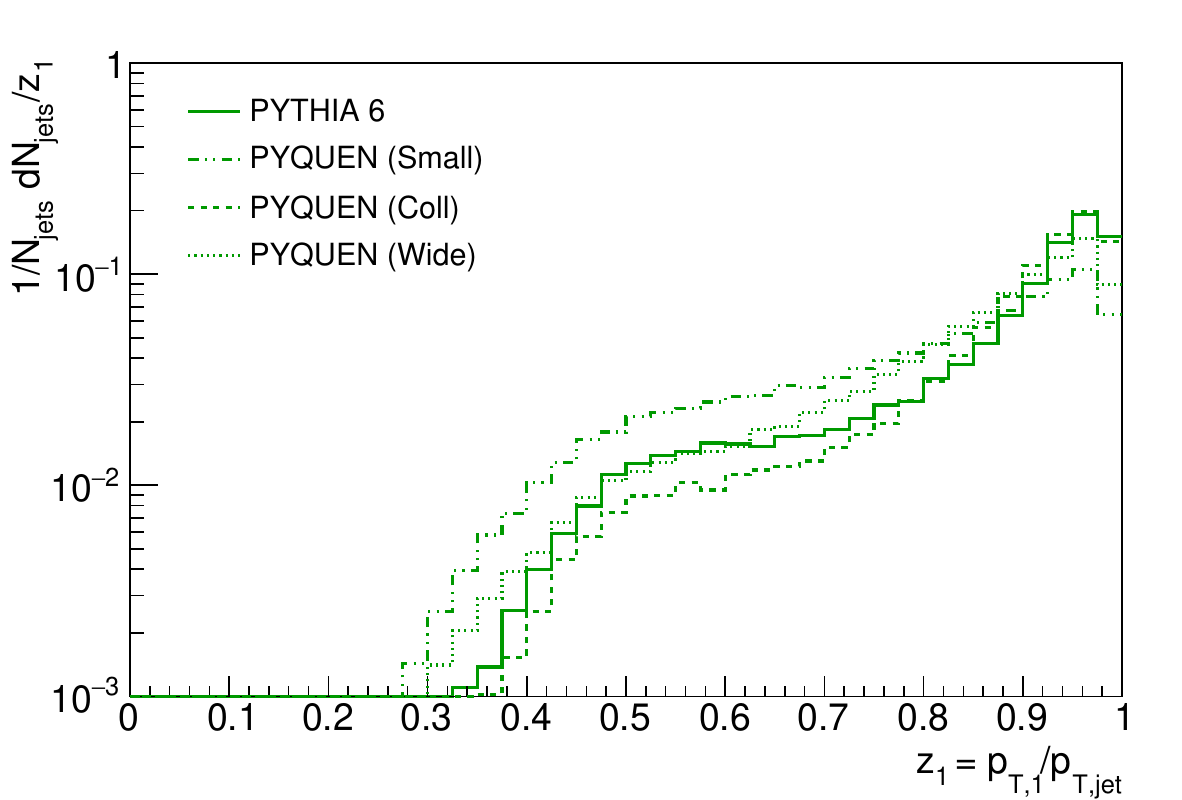}
  \caption{The fraction of transverse momenta of \akT\ $R=0.5$ jets with $\pt > 150$~\gevc\ carried by the leading \akT\ subjet reconstructed with $\rsj = 0.15$.}
  \label{fig:app1}
\end{figure}

\begin{figure}[h]
  \centering
  \includegraphics[width=0.45\textwidth, page=6]{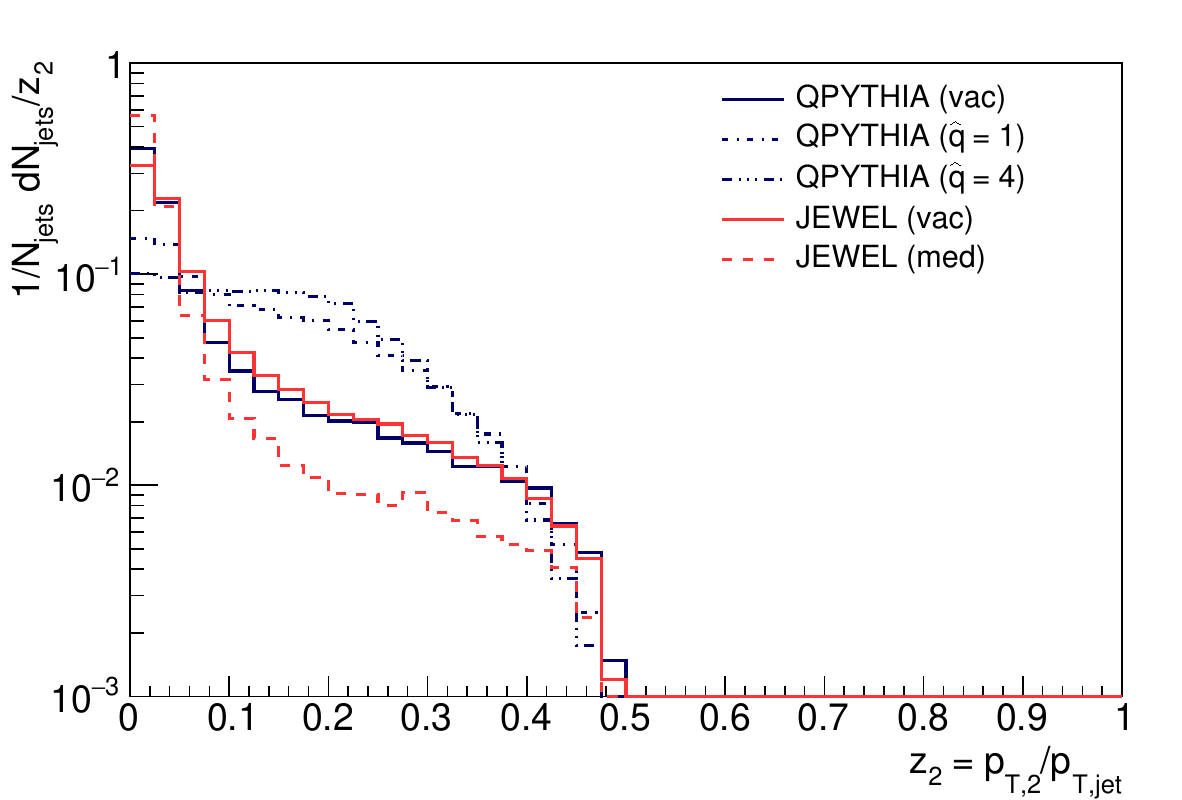}
  \includegraphics[width=0.45\textwidth, page=5]{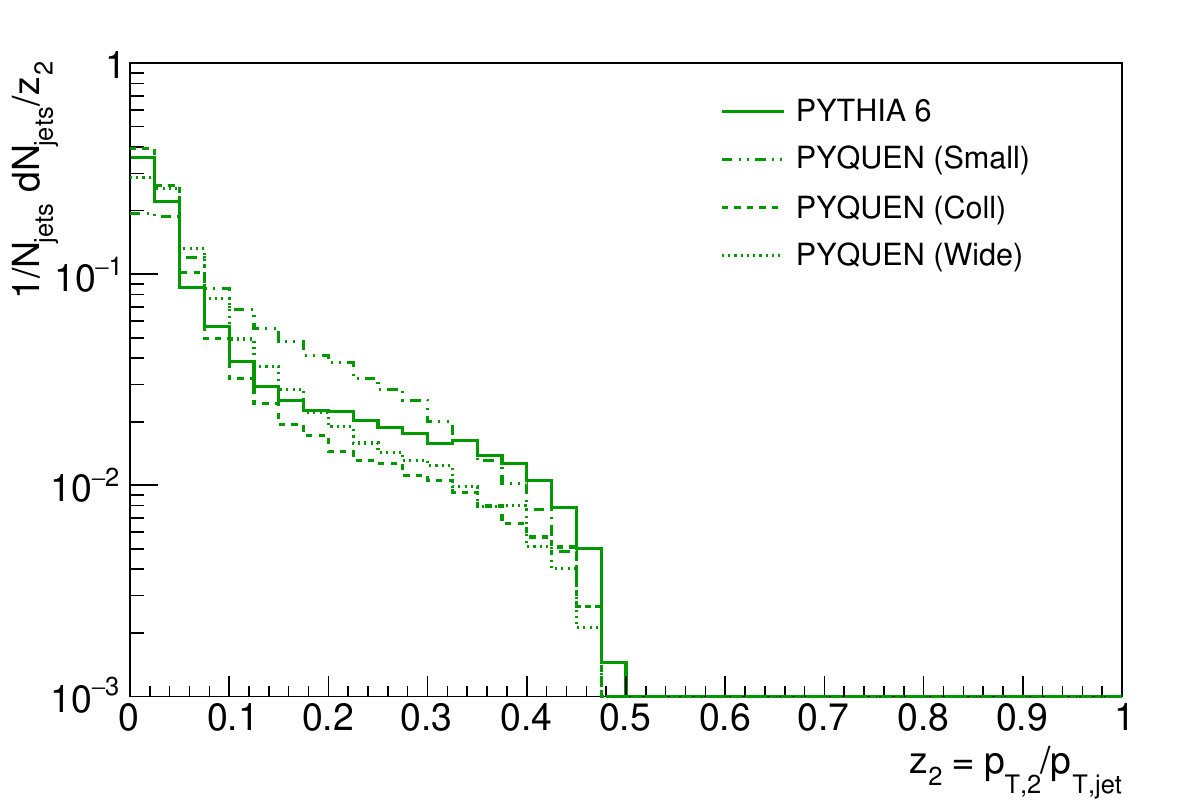}
  \caption{The fraction of transverse momenta of \akT\ $R=0.5$ jets with $\pt > 150$~\gevc\ carried by the subleading \akT\ subjet reconstructed with $\rsj = 0.15$.}
  \label{fig:app2}
\end{figure}

The gross features of the differences between the models have been discussed in terms of \dsj\ in Sec. \ref{sec:deltaS}. In this appendix we present the individual \z{i}\ distributions only for completeness and with a limited analysis.
The distribution of the fraction $z_{1}$ of the jet total transverse momentum carried by the leading subjet in jets with $\pt>150$~\gevc\ is shown in Fig.~\ref{fig:app1} for (top panel) \qpythia\ and \jewel, and (bottom panel) \pyquen\ with its three radiation pattern variants. The vacuum references  for each model --- \qpythia\ (vac),  \jewel\ (vac), and \pythia\ (for \pyquen) --- are also shown. Clearly \qpythia\ and \jewel\ modify the \z{1}\ distribution in incompatible directions. As noted for \dsj\ these observations are consistent with a collimation of jets within \jewel\ and broadening in \qpythia\ as compared to their vacuum references.
For \pyquen\ (botton panel) we find a clear separation of its different parametrisations of the angular distribution of medium induced radiation.
The  \z{2}\ distribution (the \pt\ fraction carried by the subleading subjet) shown in Fig. \ref{fig:app2} is, by definition, limited to the $0-0.5$ interval.
The differences among the \z{2}\ distributions obtained from the different models mirror those observed for $z_{1}$.
Globally, the \pt\ fraction \z{2}\ carried by the subleading subjet reflects the strongly peaked \z{1}\ distribution at large-$z$ which necessarily places the average $\z{2}$ to be below $0.1$.

\bibliographystyle{spphys}
\bibliography{template}

\end{document}